\documentclass{article}
\usepackage{graphicx}
\begin{document}
\title{On the conformally coupled scalar field quantum cosmology}
\author{Pouria Pedram\thanks{pouria.pedram@gmail.com}\\
 {\small Plasma Physics Research Center, Science and Research
Campus, Islamic Azad University, Tehran, Iran}}
\date{\today}
\maketitle \baselineskip 24pt

\begin{abstract}
We propose a new initial condition for the homogeneous and isotropic
quantum cosmology, where the source of the gravitational field is a
conformally coupled scalar field, and the maximally symmetric
hypersurfaces have positive curvature. After solving corresponding
Wheeler-DeWitt equation, we obtain exact solutions in both classical
and quantum levels. We propose appropriate initial condition for the
wave packets which results in a complete classical and quantum
correspondence. These wave packets closely follow the classical
trajectories and peak on them. We also quantify this correspondence
using de-Broglie Bohm interpretation of quantum mechanics. Using
this proposal, the quantum potential vanishes along the Bohmian
paths and the classical and Bohmian trajectories coincide with each
other. We show that the model contains singularities even at the
quantum level. Therefore, the resulting wave packets closely follow
the classical trajectories from big-bang to big-crunch.
\end{abstract}

\section{Introduction}\label{sec1}
In recent years, the question of construction and interpretation of
wave packets in quantum cosmology and its connection with classical
cosmology has attracted much attention. Moreover, some attempts have
done to construct a theory of quantum gravity and find its relation
with classical physics. Some authors use the semiclassical
approximations for the Wheeler-DeWitt (WDW) equation and consider
the oscillatory or exponentially decaying solutions in the
configuration space as classically allowed or forbidden regions,
respectively. These regions are mainly determined by the initial
conditions for the wave function of the universe. The
\emph{no-boundary} proposal  of Hartle and Hawking \cite{noboundary}
and the \emph{quantum creation} of the universe from nothingness by
Vilenkin \cite{spontaneous} are two popular proposals for the
initial conditions.

In quantum cosmology similar to the ordinary quantum mechanics, one
is usually concerned with the construction of wave functions by the
superposition of the energy eigenfunctions which would peak around
the classical trajectories and follow them in configuration space,
whenever such classical and quantum correspondence is possible
\cite{HALLIWELL,Matacz,HALLIWELL2}. However, contrary to the
ordinary quantum mechanics, an external time parameter is absent in
the theory of quantum cosmology. Therefore, the initial conditions
would have to be formulated with respect to \textit{intrinsic} time
parameter, which could be taken as the local scale factor for the
three-geometry in the case of the hyperbolic WDW equation
\cite{DeWitt}.

The construction of wave packets resulting from the solutions of the
WDW equation is a common feature of some investigations in quantum
cosmology \cite{kiefer1,tucker,sepangi,wavepacket}. To be more
specific, in Refs.~\cite{wavepacket,pedramCQG1} the construction of
wave packets in a Friedmann-Robertson-Walker (FRW) universe with a
minimally coupled self interacting scalar field is presented and
appropriate initial conditions are motivated. In
Ref.~\cite{pedramPLB3}, the authors investigated $(n +
1)$-dimensional quantum cosmology with varying speed of light and
obtained exact solutions in both classical and quantum domains. The
wave packets are constructed in such a way that they completely
correspond to their unique classical counterparts. In
Ref.~\cite{pedramJCAP}, a class of spherically symmetric Stephani
cosmological models with a minimally coupled scalar field is studied
in both classical and quantum levels \cite{pedramPLB,pedramCQG2}.
The aim of these investigations has been to find wave packets whose
probability distributions coincide with the classical paths obtained
in classical cosmology. This means that the wave packet should be
centered around the classical path and its crest should closely
follow the classical trajectory.

Here, we study a minisuperspace model describing a closed FRW
universe with vanishing cosmological constant and containing a
conformally coupled scalar field. The issue of the conformally
coupled scalar field quantum cosmology has been in vogue for more
than twenty years and basic results have been found by several
authors
\cite{HALLIWELL,Matacz,Schmidt,Laflamme,HALLIWELL3,Page2,Page,HALLIWELL4,Garay,Kim,Visser,kiefer2,barbosa,barros}.
In particular, Page \cite{Page} has also studied this model and
after solving the resulting WDW equation he obtained the solutions
for the positive, negative and zero curvatures. For the case of
positive curvature, the WDW equation casts into a
oscillator-ghost-oscillator differential equation with well-known
solutions. Moreover, Barbosa \cite{barbosa} has used one or two of
these solutions to construct the wave packets and found the Bohmian
trajectories via de-Broglie Bohm interpretation of quantum
mechanics. Therefore, in contrast to the classical scenario, he
found some non-singular solutions. In fact, the Bohmian trajectories
highly depend on the wave function of the system and various linear
combinations of eigenfunctions lead to different Bohmian
trajectories. On the other hand, since the underlying WDW equation
is second-order hyperbolic functional differential equation, we are
free to choose the initial wave function and the initial slope of
the wave function. But classically, we have a unique solution.
Therefore, we encounter with a meaningful question: Is it possible
to choose the initial condition in such a way that the resulting
wave packet completely corresponds to its unique classical
counterpart? In the previous investigations, using appropriate
initial conditions (expansion coefficients), we could construct the
wave packets which completely simulate their classical counterparts
\cite{pedramCQG1,pedramPLB3,pedramJCAP}. In particular, in
Ref.~\cite{pedramJCAP}, we obtained the wave packets so that the
quantum potential vanishes along the Bohmian trajectories. In fact,
the behavior of the quantum solution is strictly dependent to the
initial conditions imposed on the wave function. Here, at the
classical domain, the model is singular and we try to design wave
packets in such a way that they follow as closely as possible the
classical trajectories from big-bang to big-crunch. To achieve this
purpose, we propose a specific relation between the expansion
coefficients or the initial wave function and its initial slope.
Moreover, using WKB approximation we will show that the behavior of
these wave packets is in agreement with the classical motion.

The paper is organized as follows: in Sec.~\ref{sec2}, we present
the action of a homogeneous and isotropic quantum cosmology, where
the source of the gravitational field is a conformally coupled
scalar field, and the maximally symmetric hypersurfaces have
positive curvature. In Sec.~\ref{sec3}, we quantize the model and
obtain the exact solutions of the corresponding WDW equation. We
then construct the wave packets using the proposed initial
condition. Moreover, using de-Broglie Bohm interpretation of quantum
mechanics, we quantify the classical and quantum correspondence and
justify the initial condition. In Sec.~\ref{sec5}, we state our
conclusions.

\section{The model}\label{sec2}
Let us start from the Einstein-Hilbert action for the gravity plus a
conformally coupled scalar field
\begin{equation}\label{action}
S=\int d^4x \sqrt{-g} \left[\frac{1}{16\pi G}{\cal
R}-\frac{1}{2}\nabla_{\mu}\phi\nabla^{\mu}\phi-\frac{1}{12}{\cal
R}\phi^2 \right],
\end{equation}
where $g_{\mu\nu}$ is the four metric, $g$ is its determinant,
${\cal R}$ is the scalar curvature and $\phi$ is the scalar field.
Units are chosen such that $\hbar=c=1$. Here, we consider a
minisuperspace FRW model with the constant positive curvature and a
homogeneous scalar field as
\begin{equation}\label{metric}
\begin{array}[l]{l}
ds^2=-N^2(t)dt^2+a^2(t) \left[\frac{\displaystyle dr^2}{\displaystyle 1-r^2}+r^2(d\theta^2+\sin^2\theta\,d\phi^2) \right], \\ \\
\phi=\phi(t).
\end{array}\displaystyle
\end{equation}
Now, we substitute Eq.~(\ref{metric}) in Eq.~(\ref{action}) and use
the variable $\chi=al_p\phi/\sqrt{2}$, where $l_p$ is the Planck
length ($8\pi G=3l^2_p$). After discarding total time derivatives
and integrating out the spatial degrees of freedom, we obtain the
following action \cite{Page,barbosa}
\begin{equation}
S=\int dt\left[Na-\frac{a\dot{a}^{2}}{N}+\frac{a\dot{\chi}^{2}}{N}
-N\frac{\chi^{2}}{a}\right]. \label{action2}
\end{equation}
This action results in the following Hamiltonian
\begin{equation}
H=N\left[-\frac{P_{a}^{2}}{4a}+\frac{P_{\chi}^{2}}{4a}-a+\frac{\chi^{2}}
{a}\right]=N\mathcal{H}, \label{hamiltonian}
\end{equation}
where $P_{a}=-\frac{\displaystyle2a\dot{a}}{\displaystyle N}$ and
$P_{\chi}=\frac{\displaystyle2a\dot{\chi}}{\displaystyle N}$ are the
canonical momenta conjugate to $a$ and $\chi$, respectively. These
variables also satisfy the following Poisson brackets
\begin{equation}\label{poisson}
\left\{
  \begin{array}{ll}
    \{a,\chi\}=0, & \{P_a,P_{\chi}\}=0, \\ \\
    \{a,P_a\}=1, & \{\chi,P_{\chi}\}=1.
  \end{array}\displaystyle
\right.
\end{equation}
Using the above commutation relations and Hamiltonian  equation
(\ref{hamiltonian}), we find the equations of motion for the scale
factor and the scalar field
\begin{equation}\label{eqm}
\left\{
  \begin{array}[l]{l}
    \dot{a}=\{a,H\}=-NP_a/(2a), \\ \\
    \dot{P_a}=\{P_a,H\}=2N, \\ \\
\dot{\chi}=\{\chi,H\}=NP_{\chi}/(2a), \\ \\
\dot{P_{\chi}}=\{P_{\chi},H\}=-2N\chi/a.
  \end{array}\displaystyle
\right.
\end{equation}
By choosing the gauge $N=a$, we obtain the following parametric
solutions for the system
\begin{equation}\label{classicsol}
  \begin{array}{ll}
    a(t)=D \sin(t), \\ \\
    \chi(t)=D \cos(t-\theta_0),
  \end{array}\displaystyle
\end{equation}
where $D$ and $\theta_0$ are constants. Also, we have used the zero
energy condition for the super-Hamiltonian ${\cal H}=0$. It is
obvious that above solutions represent Lissajous ellipsis which are
singular in the present ($t=0$) and in the future ($t=\pi$).

\section{Quantum cosmology and wave packets}\label{sec3}
Let us study the quantum cosmology aspects of the model presented
above. The Hamiltonian can be obtained upon quantization procedure
$P_a\rightarrow -i\frac{\displaystyle\partial}{\displaystyle\partial
a}$ and $P_{\chi}\rightarrow
-i\frac{\displaystyle\partial}{\displaystyle\partial \chi}$.
Therefore, one arrives at the WDW equation describing the
corresponding quantum cosmology \cite{Page2,Page,barbosa}
\begin{eqnarray}
{\cal H} \Psi(\chi,a)=\left\{\frac{\partial^{2}}{\partial \chi^{2}}-\frac{\partial^{2}}{\partial
a^{2}}-\chi^{2}+a^{2}
\right\}\Psi(\chi,a)=0, \label{WDW}
\end{eqnarray}
where we have used a particular choice of factor  ordering and for
simplicity we have absorbed a factor of $\sqrt{2}$ into variables.
It is worth to mention that in the previous investigations
\cite{pedramCQG1,pedramPLB3,pedramJCAP}, we have also encountered
with the same form of WDW equation. There, we have constructed the
wave packets in such a way that they corresponded to the classical
cases where $\theta_0=0$. As we shall see, the usage of the real
coefficients for the even terms and the imaginary coefficients for
odd terms in the expansion, results in the symmetric solutions about
$a$ axis ($\theta_0=0$). Here, in order to study the non-symmetric
cases ($\theta_0\ne0$), we need to choose appropriate complex
expansion coefficients for both even and odd terms.

The WDW equation (\ref{WDW}) is separable in the minisuperspace
variables and solutions can be written as
\begin{equation}
\Psi_n(\chi,a)=\psi_{n}(\chi)\psi_{n}(a), \label{4.4}
\end{equation}
where
\begin{eqnarray}
\psi_{n}(z)=\left(\frac{1}{\pi}\right)^{1/4}\left[\frac{H_{n}(
z)}{\sqrt{2^{n}n!}}\right]e^{ - z^{2}/2}. \label{4.5}
\end{eqnarray}
In this expression, $H_{n}(z)$ is the Hermite polynomial and the
orthonormality and completeness of the basis functions follow from
those of the Hermite polynomials. The general wave packet which
satisfies the WDW equation (\ref{WDW}) can be written as
\begin{equation}\label{psi}
\Psi(\chi,a)=\sum_{n=\mbox{\footnotesize{even}}} A_n
\psi_n(\chi)\psi_n(a)+i\sum_{n=\mbox{\footnotesize{odd}}} B_n
\psi_n(\chi)\psi_n(a).
\end{equation}
Since the potential in each direction is an even function, the
eigenfunctions are separated in both even and odd categories. Now,
the initial wave function and its initial derivative take the form
\begin{eqnarray}
\Psi(\chi,0)=\sum_{n=\mbox{\footnotesize{even}}} A_n
\psi_n(\chi)\psi_n(0),\\
\frac{\partial\Psi(\chi,a)}{\partial
a}\bigg|_{a=0}=i\sum_{n=\mbox{\footnotesize{odd}}} B_n
\psi_n(\chi)\psi'_n(0).
\end{eqnarray}
Therefore, the coefficients $A_n$ determine the initial wave
function and the coefficients $B_n$  determine the initial
derivative of the wave function. Since the underling WDW equation
(\ref{WDW}) is second-order hyperbolic functional differential
equation, $A_n$'s and $B_n$'s are arbitrary and independent
variables. On the other hand, if we are interested to construct the
wave packets which simulate the classical behavior with known
classical positions and momentums, all of these coefficients will
not be independent. It is obvious that the presence of odd functions
of $a$ dose not have any effect on the form of the initial wave
function but they are responsible for the slope of the wave function
at $a=0$, and vice versa for even functions. For studying the
initial condition, let us consider the differential equation for the
small values of the scale factor. Near $a=0$, the WDW equation
(\ref{WDW}) takes the form
\begin{eqnarray}
\left\{\frac{\partial^2}{\partial \chi^2}-\frac{\partial^2} {\partial
a^2}- \chi^{2}\right\}\psi(\chi,a)=0. \label{eq10nearv0}
\end{eqnarray}
This PDE is also separable in $\chi$ and $a$ variables, so we can
write the solutions as
\begin{equation}\label{psi-separated}
\psi_n(\chi,a)=\psi_n(\chi)\xi_n(a).
\end{equation}
By substituting this expression in equation (\ref{eq10nearv0}), two
ordinary differential equations can be obtained
\begin{eqnarray}
\frac{d^2\xi_n(a)}{d a^2}+E_n\xi_n(a)&=&0,
\label{eqseparated1}\\
\hspace{-0.6cm}-\frac{d^2\psi_n(\chi)}{d
\chi^2}+\chi^{2}\psi_n(\chi)&=&E_n\psi_n(\chi),\label{eqseparated2}
\end{eqnarray}
where $E_n$'s are separation constants. These equations are
Schr\"{o}dinger-like equations with $E_n$s as their energy
eigenvalues. The plane wave solutions are the exact solutions for
Eq.~(\ref{eqseparated1})
\begin{equation}\label{eqplanewave}
\xi_n(a)=\alpha_n\cos\left(\sqrt{E_n}\,\,a\right)+i\beta_n\sin\left(\sqrt{E_n}\,\,a\right),
\end{equation}
where $\alpha_n$ and $\beta_n$ are arbitrary complex numbers.
Equation (\ref{eqseparated2}) is the Schr\"odinger equation for the
simple harmonic oscillator with the well known solutions
(\ref{4.5}). The general solution to equation (\ref{eq10nearv0}) can
be written as
\begin{eqnarray}\label{psi-separated2}
\psi(\chi,a)&=&\sum_{n=\mbox{\footnotesize{even}}} A^*_n
\cos(\sqrt{E_n}a)
\psi_n(\chi)+i\sum_{n=\mbox{\footnotesize{odd}}}B^*_n\sin(\sqrt{E_n}a)
\psi_n(\chi).
\end{eqnarray}
Note that, this solution is valid only for small $a$. Now, we can
obtain the initial wave function and initial slope of the wave
function
\begin{eqnarray}
\psi(\chi,0)&=&\sum_{n=\mbox{\footnotesize{even}}}A^*_n\psi_n(\chi)\label{eqinitial1}\\
\psi'(\chi,0)&=&i\sum_{n=\mbox{\footnotesize{odd}}}B^*_n\sqrt{E_n}\psi_n(\chi),\label{eqinitial2}
\end{eqnarray}
where prime denotes the derivative with respect to the scale factor
$a$. To have a complete description of the problem, we should
specify both of these quantities. On the other hand, since we are
interested to construct the wave packet with all classical
properties, we need to assume a specific relationship between these
coefficients. The prescription is that the coefficients have the
same functional form \cite{pedramCQG1,pedramPLB3,pedramJCAP}
\textit{i.e.}
\begin{equation}\label{eqcanonicalslope}
\left\{
\begin{array}{ll}
 A^*_n=C(n)\,\,\,\,\,\, \mbox{for $n$
  even},\\ \\
 B^*_n=C(n)\,\,\,\,\,\, \mbox{for $n$
  odd},
   \end{array}\displaystyle
   \right.
\end{equation}
where $C(n)$ is a function of $n$. In terms of $A_n$s and $B_n$s we
have
\begin{equation}\label{eqcanonicalslope2}
\left\{
\begin{array}{ll}
A_n=\frac{\displaystyle
1}{\displaystyle\psi_n(0)}C(n)\,\,\,\,\,\,\mbox{for $n$
  even},\\ \\
 B_n=\frac{\displaystyle\sqrt{E_n}}{\displaystyle\psi'_n(0)}C(n)\,\,\,\,\,\, \mbox{for $n$
  odd}.
     \end{array}\displaystyle
      \right.
\end{equation}
Note that, we need to specify $C(n)$ in such a way that the initial
wave function has a desired classical description. We will see that
this choice of coefficients results in a complete classical and
quantum correspondence. Using equations (\ref{psi}) and
(\ref{eqcanonicalslope2}), we can explicitly write the form of the
wave packet
\begin{equation}\label{wave}
 \Psi(\chi,a)=\sum_{n=\mbox{\footnotesize{even}}}\frac{\displaystyle 1}{\displaystyle H_n(0)}\,C(n)\,
 \frac{\displaystyle H_n(\chi)H_n(a)}{\displaystyle
 \pi^{1/2}2^nn!}e^{-(a^2+\chi^2)/2}
 +i\sum_{n=\mbox{\footnotesize{odd}}}
\,\,\frac{\displaystyle \sqrt{2n+1}}{\displaystyle
2nH_{n-1}(0)}\,C(n)\,\frac{\displaystyle
H_n(\chi)H_n(a)}{\displaystyle \pi^{1/4}
\sqrt{2^n\,n!}}e^{-(a^2+\chi^2)/2}.
\end{equation}
Although, in principle, we need to use infinite terms to construct
the wave packet, for the studied cases a few terms are sufficient to
get a reasonable accuracy \cite{pedramCPC}. Therefore, we use $50$
terms in the above summation for all studied cases.

\begin{figure}
\centerline{\begin{tabular}{ccc}
\includegraphics[width=8cm]{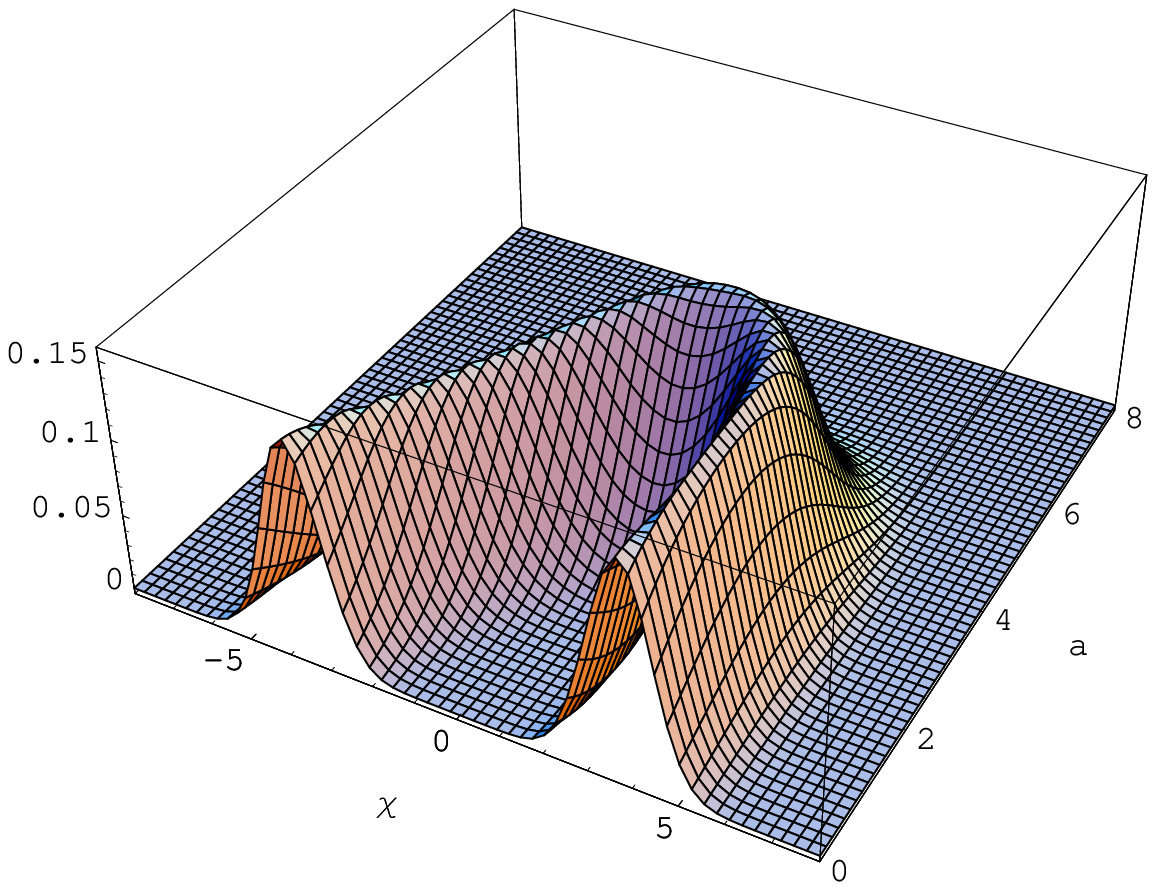}
 &\hspace{2.cm}&
\includegraphics[width=6cm]{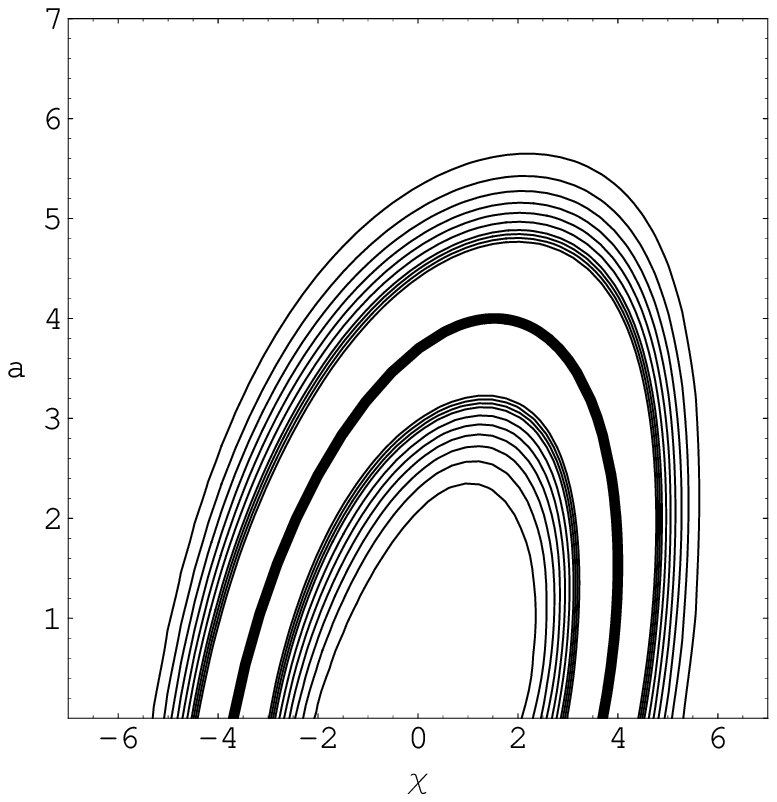}
\end{tabular}}
\caption{Left, the square of the wave packet $| \psi(\chi,a)|^2$ for
$C(n)=\frac{{\,\zeta^n}}{\,{\sqrt{2^n\,n!}}}e^{-|\zeta|^2/4}$,
$\theta_0=\pi/8$ and $|\zeta|=4$. Right, the contour plot of the
same figure with the classical path superimposed as the thick solid
line.} \label{fig1}
\end{figure}

\begin{figure}
\centerline{\begin{tabular}{ccc}
\includegraphics[width=8cm]{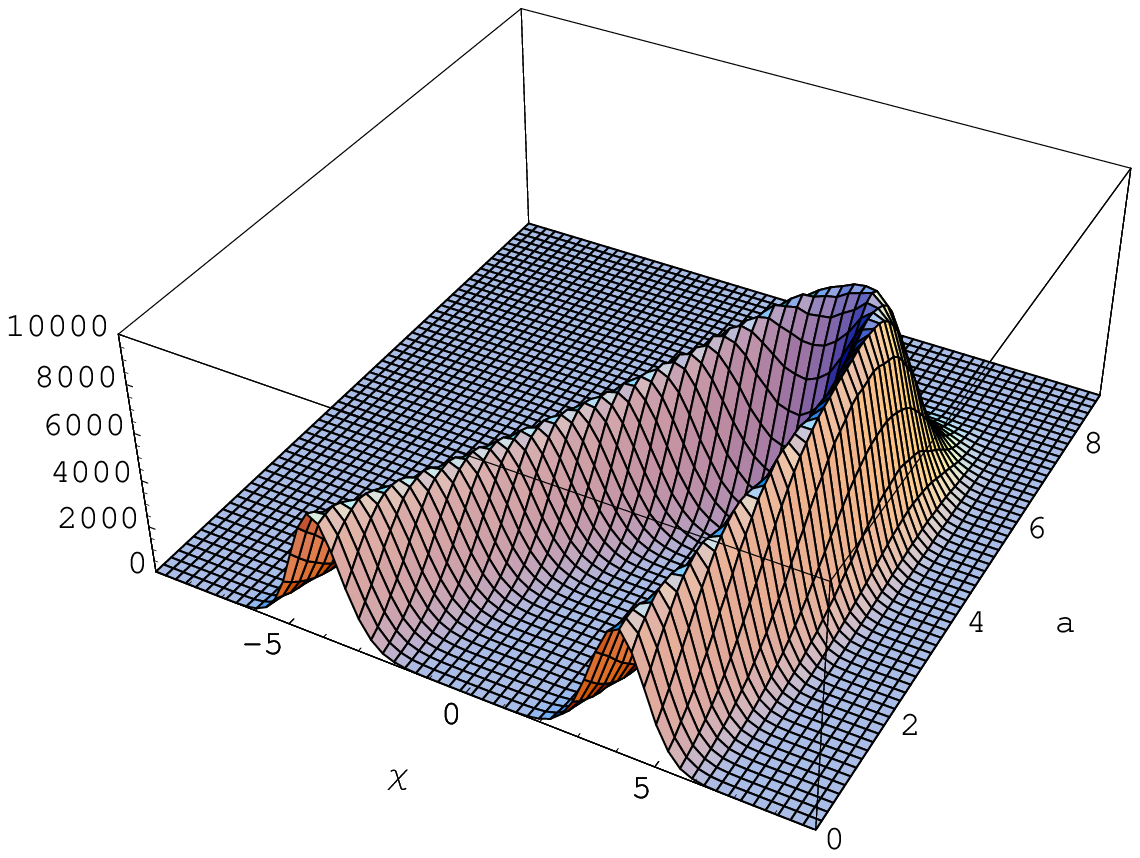}
 &\hspace{2.cm}&
\includegraphics[width=6cm]{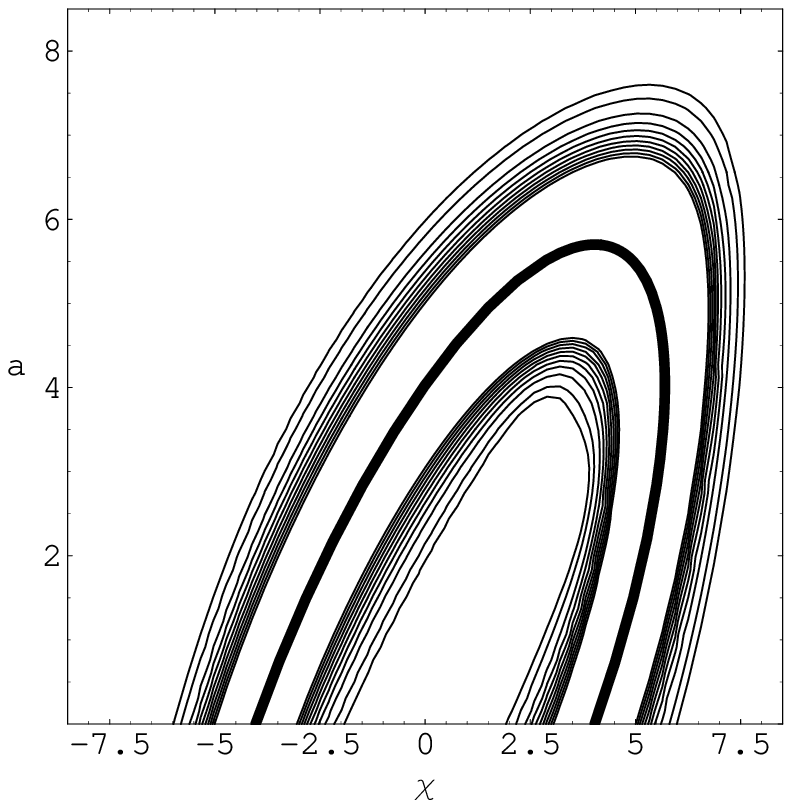}
\end{tabular}}
\caption{Left, the square of the wave packet $| \psi(\chi,a)|^2$ for
$C(n)=\frac{n^2\,\zeta^n}{\,{\sqrt{2^n\,n!}}}e^{-|\zeta|^2/4}$,
$\theta_0=\pi/4$ and $|\zeta|=5$. Right, the contour plot of the
same figure with the classical path superimposed as the thick solid
line.} \label{fig2}
\end{figure}

Figure \ref{fig1} shows the resulting wave packet for a particular
choice of initial condition
$C(n)=\frac{{\displaystyle\,\zeta^ne^{-|\zeta|^2/4}}}{\displaystyle\,{\sqrt{2^n\,n!}}}$,
where $\zeta=|\zeta|e^{-i\theta_0}$ . These coefficients are chosen
such that the initial wave function consists of two well separated
peaks. These two peaks correspond to classical initial ($t=0$) and
final ($t=\pi$) values of $\chi$, respectively. As it can be seen
from Fig.~\ref{fig1}, the square of the wave packet is smooth and
its crest follows the classical trajectory. In fact, we are free to
choose any other appropriate initial condition. For example, figure
\ref{fig2} shows the resulting wave packet with different form of
the expansion coefficients $C(n)=\frac{\displaystyle
n^2\,\zeta^ne^{-|\zeta|^2/4}}{\displaystyle\,{\sqrt{2^n\,n!}}}$.
Figures \ref{fig1} and \ref{fig2} correspond to two different
classical description with $D=4$, $\theta_0=\pi/8$ (Fig.~\ref{fig1})
and $D=5.7$, $\theta_0=\pi/4$ (Fig.~\ref{fig2}), respectively.

\begin{figure}
\centerline{\begin{tabular}{ccc}
\includegraphics[width=8cm]{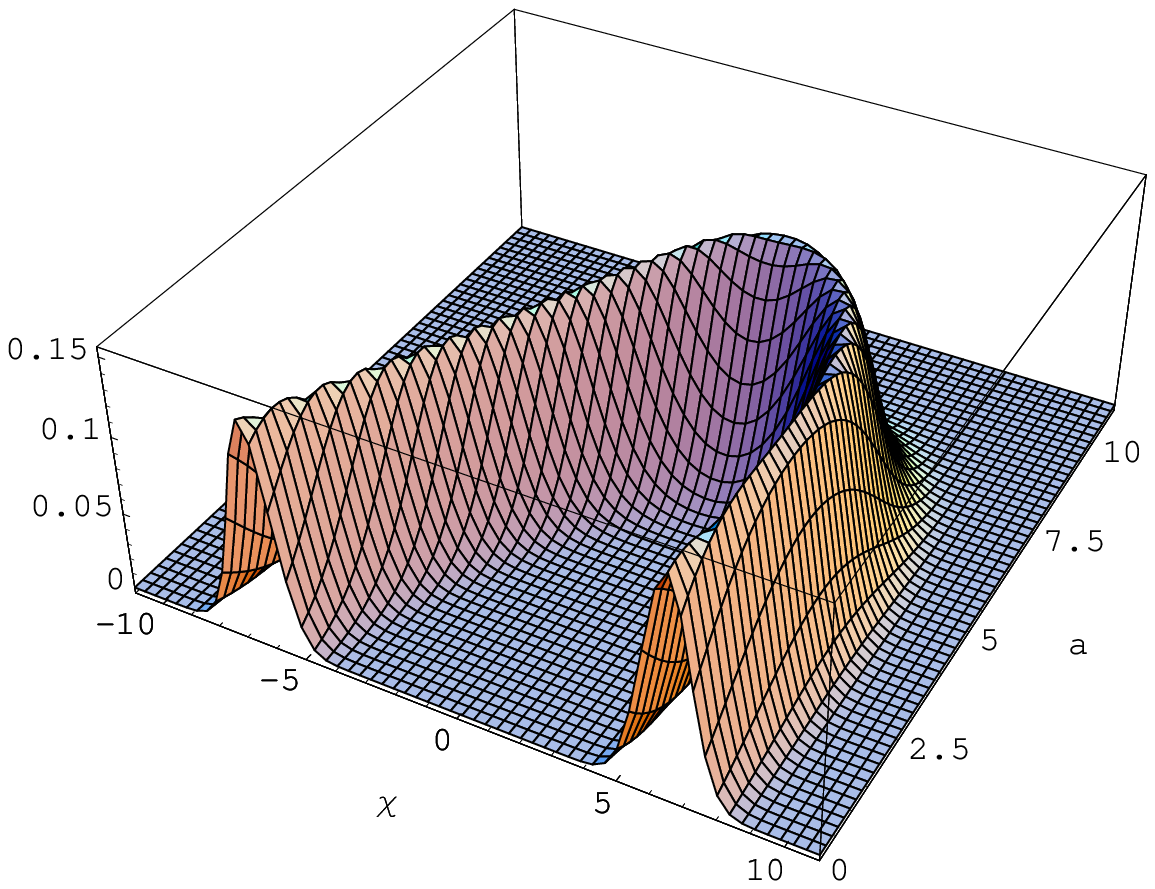}
 &\hspace{2.cm}&
\includegraphics[width=6cm]{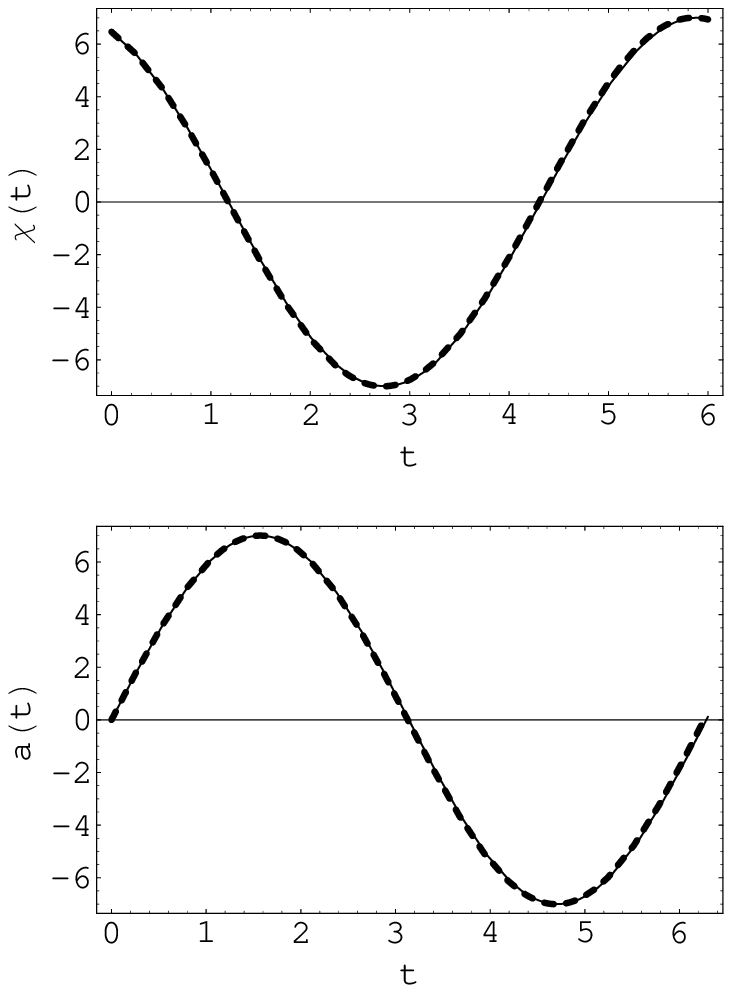}
\end{tabular}}
\caption{Left, the square of the wave packet $| \psi(\chi,a)|^2$ for
$C(n)=\frac{{\,\zeta^n}}{\,{\sqrt{2^n\,n!}}}e^{-|\zeta|^2/4}$,
$\theta_0=\pi/8$ and $|\zeta|=7$. Right, plot of $\chi(t)$ and
$a(t)$ for classical (solid line) and Bohmian (dashed line)
trajectories.} \label{figbohm}
\end{figure}

Now, let us consider the situation using the ontological
interpretation of quantum mechanics \cite{holland,nelson}. This
formalism allows us to compare quantum mechanical results with the
classical ones. Moreover, since time is absent in quantum cosmology
we can recover the notion of time using this approach. In de-Broglie
Bohm interpretation of quantum mechanics, the wave function can be
written as
\begin{equation}\label{R,S}
\Psi = R\, e^{iS},
\end{equation}
where $R=R(\chi,a)$ and $S=S(\chi,a)$ are real functions of the
scale factor and the scalar field and satisfy the following
equations
\begin{eqnarray}
\label{hje} -\frac{1}{R}\frac{\partial^2 R}{\partial \chi^2}
+\frac{1}{R}\frac{\partial^2 R}{\partial a^2}+\left(\frac{\partial
S}{\partial \chi}\right)^2-
\left(\frac{\partial S}{\partial a}\right)^2+\chi^{2}-a^{2}&=& 0,\\
\frac{\partial^2 S}{\partial \chi^2} -\frac{\partial^2 S}{\partial
a^2}+\frac{2}{R} \frac{\partial R}{\partial \chi} \frac{\partial
S}{\partial \chi}-\frac{2}{R} \frac{\partial R}{\partial a}
\frac{\partial S}{\partial a}&=&0.
\end{eqnarray}
Because of the form of the wave packet (\ref{psi}), it is more
appropriate to separate the real and imaginary parts of the wave
packet
\begin{equation}
\Psi (\chi,a)=x (\chi,a)+i y (\chi,a),
\end{equation}
where $x,y$ are real functions of $\chi$ and $a$
\begin{eqnarray}\nonumber
x (\chi,a)&=& \sum_{n=\mbox{\footnotesize{even}}}\frac{\displaystyle
1}{\displaystyle H_n(0)}\mathrm{Re}[C(n)]\frac{\displaystyle
H_n(\chi)H_n(a)}{\displaystyle
 \pi^{1/2}2^nn!}e^{-(a^2+\chi^2)/2}\\\label{xx} &-&\sum_{n=\mbox{\footnotesize{odd}}}
\frac{\displaystyle \sqrt{2n+1}}{\displaystyle
2nH_{n-1}(0)}\,\mathrm{Im}[C(n)]\frac{\displaystyle
H_n(\chi)H_n(a)}{\displaystyle \pi^{1/4}
\sqrt{2^n\,n!}}e^{-(a^2+\chi^2)/2},\\\nonumber
 y (\chi,a)&=&\sum_{n=\mbox{\footnotesize{odd}}}
\hspace{-.2cm}\frac{\displaystyle \sqrt{2n+1}}{\displaystyle
2nH_{n-1}(0)}\mathrm{Re}[C(n)]\frac{\displaystyle
H_n(\chi)H_n(a)}{\displaystyle \pi^{1/4}
\sqrt{2^n\,n!}}e^{-(a^2+\chi^2)/2}\\\label{yy}
&+&\sum_{n=\mbox{\footnotesize{even}}}\frac{\displaystyle
1}{\displaystyle H_n(0)}\mathrm{Im}[C(n)]\,\frac{\displaystyle
H_n(\chi)H_n(a)}{\displaystyle
 \pi^{1/2}2^nn!}e^{-(a^2+\chi^2)/2}.
\end{eqnarray}
These variables are related to $R$ and $S$ through equation
(\ref{R,S}) as
\begin{eqnarray}
R&=&\sqrt{x^2+y^2},\\
S&=&\arctan\left(\frac{y}{x}\right).
\end{eqnarray}
On the other hand, the Bohmian trajectories which determine the
behavior of the scale factor and the scalar field, are governed by
\begin{eqnarray}
P_{\chi} = \frac{\partial S}{\partial \chi},\\
P_a = \frac{\partial S}{\partial a},
\end{eqnarray}
where the momenta correspond to the related Lagrangian
(\ref{action2}) are $P_{\chi}=2\dot{\chi}$ and $P_a=2\dot{a}$, in
the gauge $N=a$. Therefore, the equations of motion take the form
\begin{eqnarray}
2\dot{\chi}&=& \frac{ \displaystyle x\frac{\partial y}{\partial \chi}-y\frac{\partial x}{\partial \chi}}{x^2+y^2}, \\
2\dot{a}&=&-\frac{ \displaystyle x\frac{\partial y}{\partial
a}-y\frac{\partial x}{\partial a}}{x^2+y^2}.
\end{eqnarray}
Using the explicit form of the wave packet (\ref{xx},\ref{yy}),
these differential equations can be solved numerically to find the
time evolution of $\chi$ and $a$. In the left part of
Fig.~\ref{figbohm}, we have shown the square of the wave packet $|
\psi(\chi,a)|^2$ for
$C(n)=\frac{{\,\zeta^n}}{\,{\sqrt{2^n\,n!}}}e^{-|\zeta|^2/4}$,
$\theta_0=\pi/8$ and $|\zeta|=7$. In the right part of this figure,
we have depicted $\chi(t)$ and $a(t)$ for classical (solid line) and
Bohmian (dashed line) trajectories. In fact, the obtained Bohmian
quantities versus time ($\chi(t)$ and $a(t)$) coincide well with
their classical counterparts. This shows the suppression of the
quantum potential along the trajectory due to the coincidence
between classical and Bohmian results \cite{pedramJCAP}.

\begin{figure}
\centering
\includegraphics[width=7.cm]{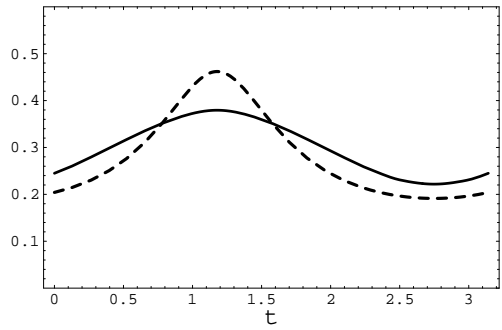}
\caption{The inverse of classical momentum $P^{-1}$ (dashed line)
and the square of the wave packet $| \psi(\chi,a)|^2$ along the
classical trajectory (solid line) for
$C(n)=\frac{{\,\zeta^n}}{\,{\sqrt{2^n\,n!}}}e^{-|\zeta|^2/4}$,
$\theta_0=\pi/4$ and $|\zeta|=4$ versus $t$.} \label{fig3}
\end{figure}

For the case $\theta_0\ne0$, the  momentum is not constant along the
classical trajectory $P=2 D\sqrt{\sin^2(t)+\cos^2(t-\theta_0)}$. In
fact, we can also obtain some information about the classical
momentum from the shape of the wave packet. For this purpose, we use
WKB approximation. In semiclassical approximation, the square of the
wave function up to the first order is related to the momentum as
\begin{eqnarray}\label{WKB}
|\psi|^2\propto\frac{1}{\displaystyle P}.
\end{eqnarray}
This equation has a simple interpretation. The low momentum
corresponds to the high probability density and vice-versa. Figure
\ref{fig3} shows the square of the wave packet $|\psi(\chi,a)|^2$
along the classical (Bohmian) trajectory which is parameterized with
time and the inverse of the classical momentum $P^{-1}$ versus time.
As it can be seen from the figure, the height of the crest of the
wave packet qualitatively shows the variation of the classical
momentum along the trajectory. The low consistency between these two
quantities is due to the approximate nature of equation (\ref{WKB}).

\section{Conclusions}\label{sec5}
We have studied a closed homogeneous and isotropic quantum cosmology
model in the presence of a conformally coupled scalar field. We have
proposed a new initial condition which results in a complete
classical and quantum correspondence. In fact, since the WDW
equation is second-order differential equation, we are free to
choose the initial wave function and the initial derivative of the
wave function. This means that we are also free to choose the
expansion coefficients. Since we are interested to have a
consistency between classical and quantum solutions, we need to
impose a particular relation between the coefficients. In this
article, we proposed a particular relation between even and odd
expansion coefficients. These coefficients determine the initial
wave function and its initial derivative, respectively. In other
words, this proposal defines a connection between position and
momentum distributions with two properties at the same time. First,
they correspond to their classical quantities and second, they
respect the uncertainty relation. To quantify the classical and
quantum correspondence, we have also studied the situation using
de-Broglie Bohm interpretation of quantum mechanics. In fact, the
Bohmian trajectories highly depend on the choice of the expansion
coefficients and various linear combinations of eigenfunctions lead
to different Bohmian trajectories. Therefore, although the
inconsistency between classical and Bohmian solutions is natural in
most cases, the quantum scenario is not always different from the
classical scenario. In this paper, we have tried to construct the
wave packets which peak around the classical trajectories and
simulate their behavior. We showed that Bohmian positions and
momenta coincide completely with their classical counterparts upon
choosing arbitrary but appropriate initial conditions. Moreover,
using WKB approximation, we qualitatively obtained the classical
variation of momentum along the path without utilizing the classical
equations of motion.

\end{document}